\documentclass[aps,twocolumn]{revtex4-1}
\usepackage{newtxtext}
\usepackage{natbib}
\usepackage[T1]{fontenc}
\usepackage{ae,aecompl}
\usepackage{graphicx}
\usepackage{times,bbm,amsmath,amssymb}
\usepackage{amsmath}
\usepackage{amssymb}
\usepackage{bm}
\usepackage{epsfig,color}
\usepackage{xcolor}
\usepackage{hyperref}
\usepackage{float,siunitx}
\usepackage[caption = false]{subfig}
\usepackage{thumbpdf,enumerate}
\usepackage{sidecap}
\usepackage[scaled=.8]{couriers}
\usepackage{pstricks}
\usepackage{multirow}
\usepackage{placeins}
\usepackage{pst-grad,bm}
\usepackage{epigraph}
\usepackage{gensymb}
\usepackage{soul}
\usepackage{acronym}
\usepackage{array}
\usepackage{xcolor,colortbl}
\usepackage{geometry}
\usepackage{booktabs, tabularx}
\usepackage{numprint}
\usepackage{scalerel}
\usepackage{txfonts}
\usepackage{tabularx}
\newcommand{\GaSPH}{\texttt{GaSPH}}

\begin{document}

\title{
{\footnotesize{\fontfamily{qcr}\selectfont
Submitted to ApSS\\
Submitted Jul. 14, 2020;\\
Accepted Oct. 17, 2020;\\
}}
\vspace{1cm}
Dynamical evolution of a young planetary system: \\
stellar flybys in co-planar orbital configuration} 

\author{Raffaele Stefano Cattolico}
\affiliation{Dep. of Physics, Sapienza, Univ. of Rome, P.le A. Moro 5, I00185 Rome, Italy}
\author{Roberto Capuzzo-Dolcetta}
\affiliation{Dep. of Physics, Sapienza, Univ. of Rome, P.le A. Moro 5, I00185 Rome, Italy}

\begin{abstract}

\begin{center}
    \normalsize{ \textbf{Abstract}}
\end{center}
Stellar flybys in star clusters may perturb the evolution of young planetary systems in terms of disk truncation, planetary migration and planetary mass accretion. We investigate the feedback of a young planetary system during a single close stellar encounter in a typical open young stellar cluster. We consider 5 masses for the stellar perturbers: 0.5, 0.8, 1, 3 and 8 M$_{\odot}$, in coplanar, prograde and retrograde orbits respect to the planetary disk, varying the pertruber-host star orbital periastron from 100 AU to 500 AU. We have made 3D modelizations with the smooth particle hydrodynamics code {\GaSPH} of a system composed by a solar type star surrounded by a low density disk where a giant planet is embedded in. We focus on the dynamical evolution of global parameters characterizing the disk and the planet, like the Lagrangian radius containing the $63.2\%$ of the mass of the disk, the distance of the planet to its host star, the planet orbital eccentricity and the planetary mass accretion. We find that the most part of the simulated systems show a significant disk truncation after a single close encounter, a final orbital distance of the Jovian, from the central star, lower than the unperturbed case and, finally, the perturbed systems show a final mass accretion of the Jovian planet larger than the non-perturbed case.
Therefore, stellar flybys significantly perturb the dynamics of a young planetary system, regardless the orbital configuration of the stellar perturber. In such experiments, the final disk radius and the orbital parameters of the Jovian planet are considerably affected by the stellar close encounter.

\vspace{5mm}

\end{abstract}

\maketitle

\section{Introduction}
\label{introduction}
Stars tend to form in stellar aggregates or clusters (\cite{Hillenbrand}; \cite{Palla}), even our Sun might have formed this way (\cite{Dukes}).
Consequently, the planetary architectures around stars have likely been affected by several stellar gravitational interactions (\cite{Picogna}, \cite{Rosotti}, \cite{Cuello}). 
If the planetary system is aged in the interval 3 - 12 Myr, it, likely, still shows a circumstellar disk (\cite{Haisch}). Moreover, a coexistence between a  forming planet and the gaseous disk is predicted both by theoretical models (\cite{Pollack}) and observations, as shown by several surveys obtained by instruments like SPHERE (\cite{Avenhaus}) through a direct imaging in the near-infrared, or in radio bands by ALMA (\cite{Andrews}). Several studies have simulated circumstellar disks using different approaches, starting from a clump of ``test particles'' to a full hydrodynamic approach. A pure $N$-body test particles approach represents only a rough approximation, although useful, of the dynamics of a real circumstellar disk because neither viscosity nor pressure forces are taken into account. 
\newline
\cite{Rosotti} considered the evolution of a circumstellar disk in a stellar cluster environment combining $N$-body simulations for the stellar cluster and a hydrodynamic code to model the gaseous disk.
However, these authors do not include any planet or protoplanet inside the circumstellar disk; moreover each disk is simulated with a very large stellar ``accretion radius'' (in the range from 1 AU to 20 AU) so that the dynamics in the inner region of the disk is not resolved.
\cite{Breslau} used an $N$-body approach to study the tidal truncation of a circumstellar disk.
Assuming prograde, coplanar encounters with a stellar perturber, they find that the final disk size is a function of the periastron of the encounter and the mass ratio between the host star and the stellar perturber.
They also estimated the portions of gas/dust material captured by the passing-by star and remaining around the primary star after the flyby, finding that prograde encounters are the most effective and destructive. By contrast, retrograde encounters were found to leave the disk almost intact.
As \cite{Rosotti}, also  \cite{Breslau} did not consider any planet or protoplanet in their simulations. The presence of a planet in a stellar disk and the mutual role after a stellar encounter was studied by \cite{Picogna}. 
They show that if the circumstellar disk is still present during a stellar flyby and it is sufficiently bound to survive the encounter, the planetary system returns, approximately, to the initial configuration on a short timescale.
Using a 3D hydrodynamic modeling of the circumstellar disk, they explored different inclinations between the disk, the planet orbit and the incoming star trajectory. Both the perturber and the central system are equipped by a low mass disk.
\newline
Our work inserts in this line of analysis of the disk+planet interaction with a perturbing star by considering a wide range of masses for the stellar perturber and different distances of minimum approach. Actually, our purpose is to simulate and investigate a statistically significant sample of perturbed planetary systems, although limiting ourselves to the case of retrograde and prograde coplanar orbits. A more general investigation is postponed to future works. \newline
The organization of the paper is the following. In section \ref{model} we introduce our model, based on the use of a new smooth particles hydrodynamics (SPH) code developed in our group, called {\GaSPH} and the initial configuration of the disk, the parent star, the revolving embedded planet and the stellar perturbers. The results of our investigation are presented in section 3 and discussed in section 4. We draw some relevant astrophysical conclusions and implications in section 5. Finally, we summarize our results in the Appendix A.


\section{The model}
\label{model}
In this work we study the consequences of the interaction of a fly-by star with a star surrounded by a gaseous disk where and embedded planet is orbiting. The dynamical evolution of the whole system is followed by means of 3D simulations using the code {\GaSPH} (\cite{Pinto}). {\GaSPH} is an optimized SPH code which consider self gravity by mean of a tree-based scheme. As it is standard in SPH, the values of a set of characteristic quantities (density $\rho$, pressure \textit{P}, specific internal energy \textit{u}, velocity $\mathbf{v}$) are calculated by means of an interpolation procedure which makes use of a proper ``kernel'' function. The central star, the revolving (giant) planet and the stellar perturber are treated as single, massive \textit{sink particles} (\cite{Bate}). Sink particles are considered as interacting with normal gas particles only via gravity: SPH particles approaching the sink particle  within a certain, given, distance, called  \textit{sink radius}, are accreted by the massive (sink) object. In our scheme, the value of the sink radius is set at the beginning of the calculation and kept fixed along the system time evolution. The distribution of the SPH particles was axisymmetric and sampled from a profile characterized by both an inner and an outer radial cutoff,  $R_{in}$ and $R_{out}$. We assumed  $R_{in}\simeq 0.015 R_{out}$ where $R_{out} = 70$ AU. 
The initial value of the outer radial cutoff distance (within which we sampled the distribution of SPH `particles') was set to 70 AU, because it is a careful compromise between the typical radius of the Minimum Solar Mass Nebula \cite{Desch} and the closest approach of our stellar perturber (100 AU). A radius smaller than 70 AU could be physically unrealistic and a value larger than 70 AU, in the case of a periastron of 100 AU of the stellar perturber, could generate computational errors due to shock waves hardly to follow with our limited resolution.
We set the disk thermal profile according to the ``flared disk'' model, for which the ratio $\frac{H}{R}$  increases with $R$, where $R$ is the radial coordinate on the disk plane and $H$ is the vertical pressure scale height $H = c_{s} / \Omega_{k}$ , where $c_{s}$ is the sound speed and $\Omega_{k}$ is the Keplerian frequency of the revolving disk (\cite{Garcia} and \cite{Dullemond}). Here a vertically \textit{isothermal} approximation is used, which assumes that any radiative energy input from the star is efficiently dissipated away: this means that the cooling time is far shorter than the dynamical timescales. The disk temperature thus follows the law:
   \begin{equation}
   T(R,z) = {T_{0}} \left({\frac{R}{R_{0}}}\right)^{-q},
   \end{equation}
   where R$_0$ represents a (radial) scale length, and $q$ is commonly taken as $q = 1/2$. T$_0$ is the  temperature at the radial (on the disk) distance $R = R_0$.
   Here we use a slightly smaller slope, $q=3/7$ as adopted by \cite{D'Alessio}, by making the assumption that the thermal processes in the inner layers of the disk do not affect its dynamical stability (see also \cite{Pinto}). 
   The surface density radial profile is:
   \begin{equation} \label{eq:1}
   {\Sigma}(R) = {{\Sigma}_{0}} \left({\frac{R}{R_{1}}}\right)^{-p} ;
   \end{equation}
   where $p= 3/2$, according to the Minimum Mass Solar Nebula model (\cite{Hayashi}). \\
   The scale radius, $R_{1}$, is set to $10$ AU \cite{Pinto}, while the value of the
   density scale ${\Sigma}_{0}$ is determined by the initial disk mass.
   
   \begin{table}
   \caption{ \label{table:set_simulations} Values of parameters characterizing the stellar flybys.}
   \begin{tabular}{r  l}
   \hline \hline
    \\
    Stellar perturber mass [M$_{\odot}$]: & 0.5, 0.8, 1, 3, 8\\
    Periastron  r$_{p}$ {[}AU{]}:              & 100, 200, 300, 400, 500\\
   Orbital inclination i$^{\circ}$:                     & 0$^{\circ}$, 180$^{\circ}$ \\
   \\
   \hline \hline
   \end{tabular}
   \end{table}

\subsection{Initial conditions}
The environment, where our disk and planet are embedded in, is thought to be that of a young open star cluster, which, typically, contains up to $10^3$ stars and has 1 pc as typical half mass radius.

For the stellar perturber, we chose five  values of the mass: $0.5 ,0.8, 1, 3, 8$ M$_{\odot}$.
These values cover sufficiently the range of star masses of an open cluster.
The speed at infinity of the stellar perturber is assumed $v_\infty = 2$ km/s, also compatible with the typical average speed in galactic open clusters. 
We studied the interaction of a flyby star moving on the same plane of the disk surrounding the primary star, on both \textit{prograde} orbits (co-rotating respect to the planet revolution around the primary) and \textit{retrograde} (counter-rotating) orbits. These 2 cases correspond to inclination angles $0^{\circ}$ and $180^{\circ}$, respectively.
For each stellar perturber mass and inclination, we selected the following periastron distances (r$_{p}$): 100, 200, 300, 400, 500 AU.
Table \ref{table:set_simulations} resumes the values of the various parameters. Starting from the given periastron distance, to provide the proper initial conditions for the star perturber motion we needed the impact parameter, $p$, of the incoming perturber, which is given by the relation \cite{Spurzem}:
\begin{equation} \label{eq:1}
p = r_{p}\sqrt{1 + 2\frac{GM_{tot}}{r_{p}{v_\infty}^{2}}}, 
\end{equation}

where $M_{tot}$ is the total mass of the system (central star+stellar perturber+disk+Jovian planet). \\
For our orbital integration a relevant role is played by the starting position of the stellar perturber, that in our investigation, was assumed at the distance at which its tidal perturbation on the star+disk+planet system has a fractional amplitude:
\[
\delta =  \frac{F_{tid}}{F_{rel}} = 10^{-4}
\]
where F$_{tid}$ and F$_{rel}$ are the initial tidal force between the planetary system and the stellar perturber, and the initial relative force between each component of the planetary system (\cite{Fregeau}).
The time extension of each simulation is assumed as twice the interval of time from the initial position of stellar perturber to the periastron of the orbit of the perturber.

\subsubsection{Disk set-up}
We simulated the gaseous disk using $2 \times 10^4$ SPH particles. We found that 
this number represents a good compromise between the computational weight and the reliability of the simulation results in the frame of a statistically significant sample of numerically simulated systems, a good compromise .
The initial mass of the disk is assumed $0.0091$ M$_{\odot}$, which is, basically, the Solar Minimum Mass Nebula ($0.01$ M$_{\odot}$) (\cite{Hayashi}) minus the embedded Jovian planet mass.
As we said above, the initial inner cut-off of the disk, R$_{in}$ is set to 1 AU and the external cut-off R$_{out}$ is 70 AU.\\
At the scope of a quantitative analysis of the feedback of the disk to the external perturbation, we followed \cite{Bate} and consider as ``radius'' of the SPH disk the Lagrangian radius of $63.2\%$ of its total mass.

\subsubsection{Planet set-up}
We consider a fully formed giant planet with a mass of $1$ M$_{Jup}\simeq M_\odot/1000$. 
Its initial orbit was assumed circular and coplanar to the disk, with a 10 AU radius, well beyond the typical solar-like snow line ($\sim 2.7$ AU).
We account for mass accretion on the planet, whose mass grows whenever an SPH particle enters the planetary sink radius. To simulate a realistic accretion we set, as suggested in \cite{Ayliffe}, a planetary sink radius $\frac{r_{H}}{50}$ where r$_{H}$ is the initial Hill radius \cite{Hill} of the planet:
\begin{equation} \label{eq:1}
r_{H} = a \left(\frac{M_{P}}{3M_{\star}}\right)^{1/3};
\end{equation}
where $a$ is the initial distance to the host star
of the planet with initial mass $M_{P}$ revolving around a star with initial mass $M_{\star}$.

\section{Results}
\label{results}
To measure the response of the Jovian planet to the perturbation induced by the flyby star, we consider proper averages of the planet distance from the host star and of its orbital eccentricity, as well as its mass accretion.
The value of these parameters at the end of the simulation is evaluated making a time average over an interval of $10$ times the initial planetary orbital period, which is $31.60$ yr.
As we said above, the dynamical perturbation of the disk is quantified by the final Lagrangian radius containing $63.2\%$ of the disk mass, $r_{63.2}$.

For each stellar perturbation we compare the final planetary distance from the host star, the final orbital  eccentricity, the final planetary mass accretion and the final $r_{63.2}$ of the gaseous disk to those of the corresponding unperturbed model, so to check how exogenous gravitational perturbations affect the intrinsic dynamical evolution of a young planetary system.
\begin{figure}
\centering
\caption{\label{fig:05_LR} Time evolution of the Lagrangian radius of the $63.2\%$ of the mass of the gaseous disk (green solid-line) during a stellar flyby of a perturber of $0.5$ M$_{\odot}$ and a $r_{p} = 100$ AU (whose distance to the planet hosting star is given by the black dashed line) in a prograde (a) and retrograde (b) configuration. 
With (c) we report the model without the stellar perturber.
}
\vspace{2mm}
\includegraphics[scale=0.39]{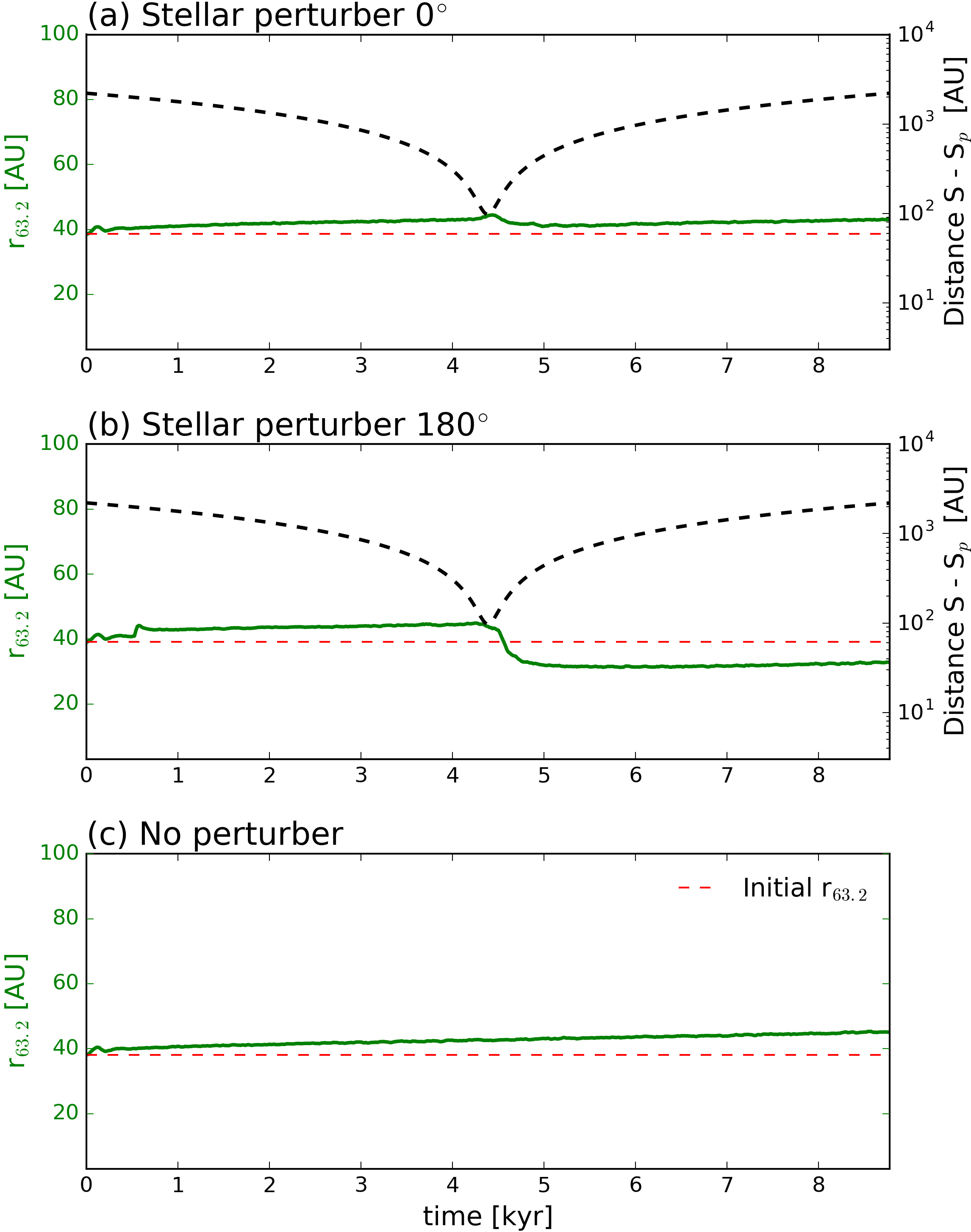}
\end{figure}
\begin{figure}
\centering
\caption{\label{fig:8_LR} As in Fig. \ref{fig:05_LR} but for a perturber of $8$ M$_{\odot}$.
}
\vspace{2mm}
\includegraphics[scale=0.38]{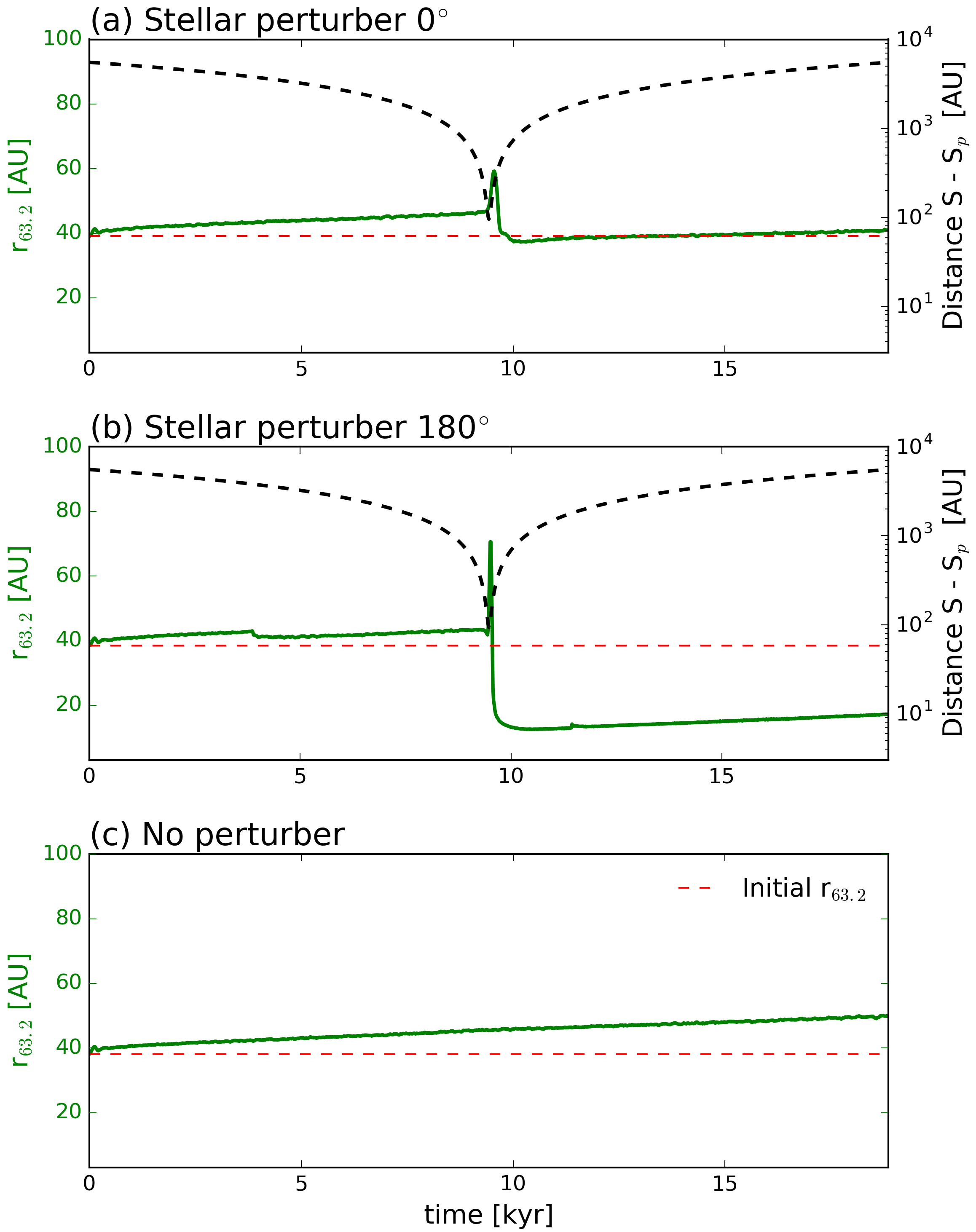}
\end{figure}
\subsection{Disk feedback}
The final $r_{63.2}$ of the disk, after both prograde and retrograde encounters, shows several cases of truncation and diffusion with respect to the analogous cases without external perturbation.
In the whole set of our experiments, the $88\%$ of the simulated systems show a disk truncation after a single stellar flyby while the remaining $12\%$ show, instead, a final disk diffusion.
When considering the case of prograde stellar perturbations, we see that the $92\%$ of the cases under study show a final disk truncation.
The same systems, but with the stellar perturber in retrograde motion, show a final disk truncation in the $80\%$ of the cases.
Although the retrograde case represents the orbital configuration with the shortest  effective interaction time between the planetary system and the stellar perturber, it shows, in the $20\%$ of the retrograde models, a more perturbative performance then the prograde ones.   
That is, for instance, the case of the model with a stellar perturber of mass $0.5$ M$_{\odot}$ and r$_{p} = 100$ AU. 
Actually (see also Fig.\ref{fig:05_LR}), its final $r_{63.2}$ is only $37.73\%$ of the final $r_{63.2}$ of the unperturbed case, and it is $18.7\%$ lower than the prograde one. 
The disk truncation occurs fast after the close passage of the stellar perturber ($100$ AU) and, later it slightly increases following an adiabatic expansion. 
In our experiments, the strongest disk truncation is in the model with the stellar perturber $8$ M$_{\odot}$ with the r$_{p}=100$ AU, in retrograde configuration.
As shown in Fig.\ref{fig:8_LR}, its final $r_{63.2}$ is $62.5\%$ lower than the final $r_{63.2}$ of the unperturbed case and it is $50.1\%$ lower than the prograde one.
In spite of the high mass of the stellar perturber and its small pericenter distance to the system, the disk is not totally disrupted but just severely truncated.
\begin{figure*}
\centering
\caption{\label{fig:a} Time evolution of the distance between the Jovian planet (P) and its host star (S) (blue solid line) during the stellar flyby of a perturber of $8$ M$_{\odot}$ (S$_{per}$) with a $r_{per}=500$ AU in retrograde configuration (whose distance to the planet hosting star is given by the black dashed line).
The magenta solid line represents the host star-planet distance in the unperturbed case.}
\vspace{2mm}
\includegraphics[scale=0.5]{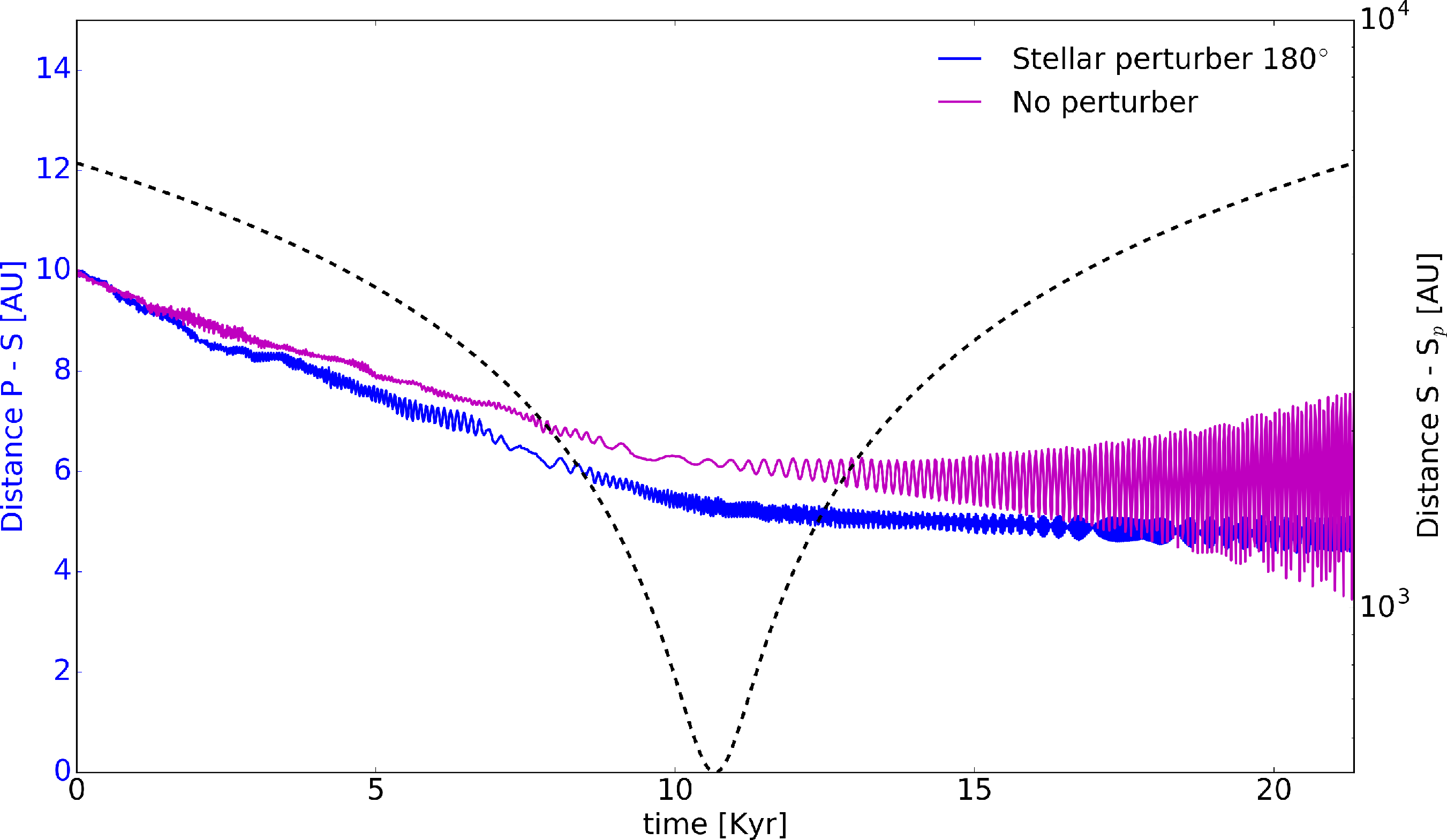}
\end{figure*}
\subsection{Jovian planet feedback}
The final (mean) distance $\langle r \rangle$ of the Jovian planet from the central star and its final (mean) orbital eccentricity $\langle e \rangle$ represent relevant orbital planetary elements for our analysis.
The $96\%$ of the studied systems show a final $\langle r \rangle$ lower than the corresponding unperturbed cases.
All the systems where the stellar perturber was on a prograde orbit experience a final decrease of $\langle r \rangle$ with respect to the unperturbed case. The same systems, but in retrograde configuration, give in $92\%$ of the cases a final decrease of $\langle r \rangle$ with respect to the unperturbed model.
This decrease of $\langle r \rangle$ is a clear planetary ``migration'' inward, in particular it is a type II planetary migration.
The longest \textit{migratory journey} in our experiments occurs in the model with the $8$ M$_{\odot}$ perturber for the r$_{p}=500$ AU, in retrograde configuration (see Fig.\ref{fig:a}): the migration is from an initial distance of $10$ AU from the host star down to $4.72$ AU over $697$ initial orbital periods.
Compared to the case without a stellar perturber (magenta solid line in Fig.\ref{fig:a}) both the systems start an initial pure planet-disk interaction but, in the model with the stellar perturber, the perturbed planet shows more ``resiliency'' than the unperturbed one. 
Considering the planetary eccentricity, the $100\%$ of the studied systems show a final eccentricity lower than that in the model without stellar perturber, regardless if the stellar perturber is on prograde or retrograde motion.
Moreover, the trajectories of the Jovian planet in the model with massive perturbers, $3$ or $8$ M$_{\odot}$, show a final eccentricity slightly greater than with a low-mass perturber ($0.5$, $0.8$ or $1$ M$_{\odot}$).
\begin{figure*}
\centering
\caption{\label{fig:ACCRETION} Time evolution of the planetary mass accretion of the Jovian planet (red solid line) during a stellar flyby of a perturber of $8$ M$_{\odot}$ (S$_p$) with a  r$_{p}$ $500$ AU, in retrograde configuration (the distance of the perturbing star to the planet hosting star is given by the black dashed line).
The magenta solid line represents the unperturbed case.}
\vspace{2mm}
\includegraphics[scale=0.18]{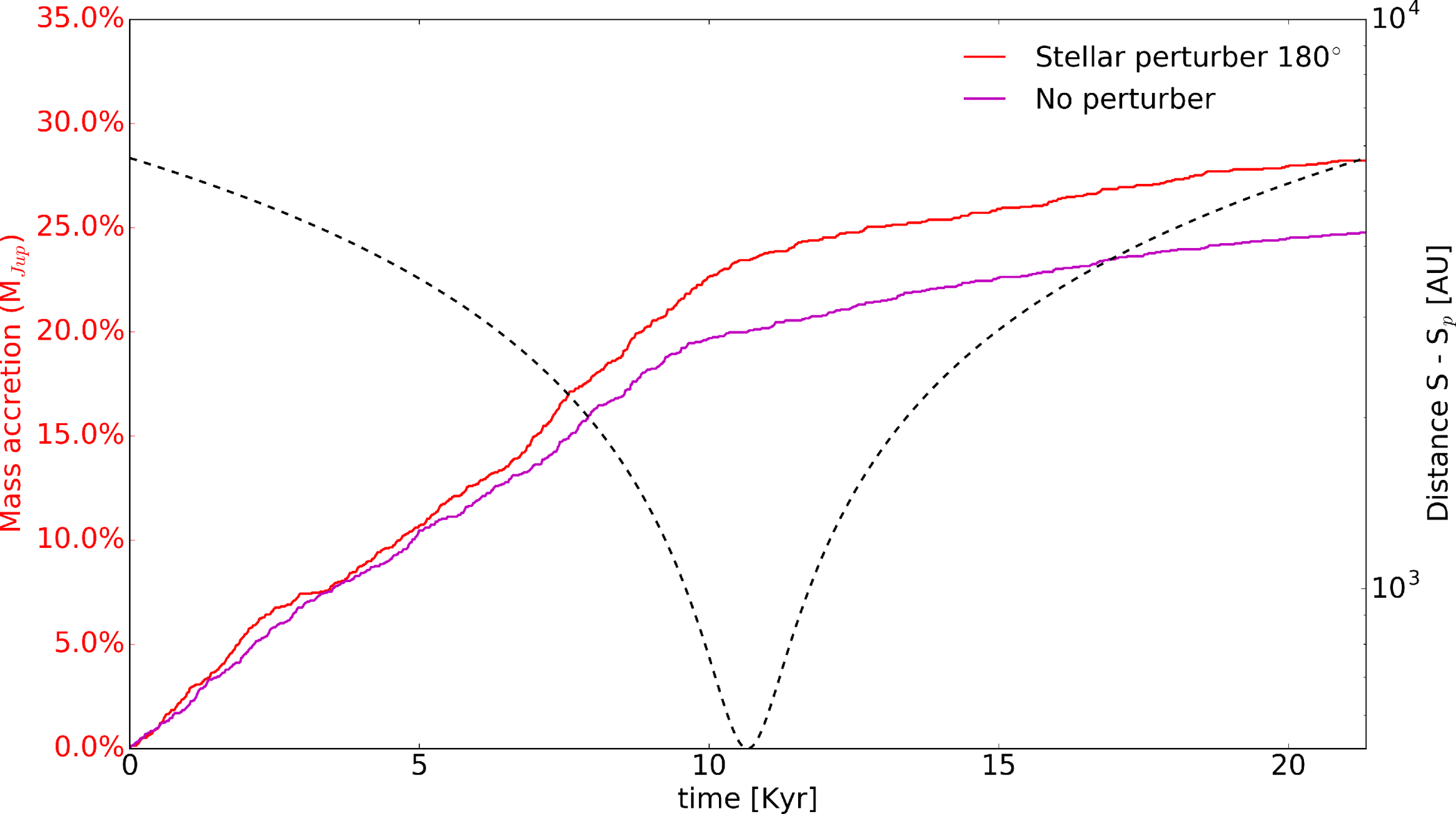}
\end{figure*}
\subsection{Planetary mass accretion}
In our experiments the $12\%$ of the perturbed systems show a final mass accretion larger than the unperturbed case independently of prograde or retrograde perturbation.
The increase of the planetary mass accretion is more evident in models with massive stellar perturbers ($3$ and $8$ M$_{\odot}$) than in the models with low-mass stellar perturber ($0.5$, $0.8$ and $1$ M$_{\odot}$). 
In the case of the $8$ M$_{\odot}$ perturber (see Fig.\ref{fig:ACCRETION}) with r$_{p} = 500$ AU we find the maximum mass planetary accretion in our experiments: $28.23\%$ of the initial planetary mass. 
Fig.\ref{fig:ACCRETION} shows two different regimes in the planetary mass accretion.
A larger slope is followed by a smaller one; the high mass accretion regime lasts until a gap is formed along the planet path in the disk. In the perturbed case the slope in this initial phase becomes greater than in the unperturbed one because of the gas compression toward the central star caused by the external perturber. 
After the closest approach of the stellar perturber, the perturbed model (red solid line in Fig.\ref{fig:ACCRETION}) shows also a planetary mass larger than the unperturbed one.
This increase is due to the partial replenishment of the gap by the gas SPH particles, increasing therefore the planet mass during its orbital revolution.
In this specific model the final planetary mass accretion is $16.61\%$ greater than the final unperturbed planetary mass accretion ($23.54\%$ of the initial mass).

\section{Discussion}
\label{discussion}
In our simulations we investigated encounters of a flyby star with a young planetary system in prograde and retrograde configurations for 5 different perturber masses ($0.5$ M$_{\odot}$, $0.8$ M$_{\odot}$, $1$ M$_{\odot}$, $3$ M$_{\odot}$, $8$ M$_{\odot}$).
Previous works (e.g. \cite{Clarke}, \cite{Dai}, \cite{Xiang-Gruess}, \cite{Picogna}) found that the final disk shapes are more affected by prograde and coplanar encounters, while they found barely perturbed disks upon retrograde encounters.
In our experiments presented in this paper we obtain that the prograde orbits are not necessarily the most perturbative ones; on the contrary, we observe that relevant perturbations are randomly led by prograde and retrograde encounters.
In  all the cases studied we see remarkable perturbations caused on the gas disk+planet system by the stellar flyby, regardless the orbital configuration of the stellar perturber.
Anyway, we do not see, after the stellar flybys, any evidence of spirals or warps as in  \cite{Cuello}, but, at most, a diffuse gaseous ``halo'' localized at the edge of the disk.
However, we note that \cite{Cuello}, \cite{Clarke},  \cite{Dai} and \cite{Xiang-Gruess} simulated a pure gaseous disks or a gaseous disk with a dusty component without presence of a planet inside the disk.
Possible explanations for the mentioned differences are both structural (our system is a more compact system due to the presence of a fully formed Jovian planet) and/or dynamical. Actually we assumed the same initial velocity for the stellar perturbers for each simulated model while in  \cite{Cuello} the initial velocity of the stellar perturber was assumed a function of the periastron r$_{p}$, inducing a sort of model dependence of the final outputs in each model simulated.
Finally, the initial phase respect to the central star of the Jovian planet might play a role in order to explain the resiliency of our perturbed gaseous disk and even the evolution of the planetary orbital parameters after the passing by star.
\newline
In a future investigation we will set different values of $v_\infty$ for the perturber, basing on proper dynamical considerations of the stellar environment and we will also vary the initial phase of the giant planet to investigate any potential phase-dependency.
The Jovian planet, initially revolving on $10$ AU circular orbit around the host star, shows a significant orbital perturbation at the end of all of our simulations, regardless the configuration of the mass and the orbit of the stellar perturber.
The overall result is that the planet moves inward the host system and this migration, compared to the model without a stellar perturber, is clearly enhanced. 
After the closest passage of the perturbing star we observe a fast migration followed by a slight increase of the planet orbital eccentricity.
Additionally, it is seen that the final eccentricity is damped more efficiently in the perturbed disk than in the isolated case. The planetary mass accretion of a Jovian planet in the perturbed system, regardless the orbital configuration of the stellar perturber, is generally smaller than in the unperturbed case: the perturbed disk slowly accretes the mass of the Jovian planet, although its faster inner migration.
However, no clear correlation is found among the final planetary parameters and the orbital parameters of the stellar perturber.

\section{Conclusions}
\label{conclusions}
In this work we studied the dynamical evolution of a young planetary system affected by a single 
stellar flyby as it would happen with some frequency in a typical open star cluster.
We consider a system composed by a central star of 1 M$_{\odot}$, surrounded by a gaseous disk of mass $0.009$ M$_{\odot}$ containing a completely embedded Jovian planet set on an initial $10$ AU radius circular orbit.
We chose 5 stellar perturber  masses (0.5 M$_{\odot}$, 0.8 M$_{\odot}$, 1 M$_{\odot}$, 3 M$_{\odot}$ and 8 M$_{\odot}$).
For each perturber mass, we changed the distance of the closest approach from $100$ AU up to $500$ AU. 
We simulated $50$ stellar flybys in coplanar  prograde and retrograde orbital configurations.
We quantify the final disk size, the inward migration of the Jovian planet, the damping of its eccentricity and its mass accretion.
All the systems remain globally bound after the stellar flyby and we do not see any clear correlations between the final orbital parameters of the Jovian planet and the orbital parameters of the stellar perturber.
Finally, the most part of our simulated systems experienced a clear inward planetary migration faster than the unperturbed case.


\bibliographystyle{mnras}

\newpage

\appendix

\begin{table*}
\centering
\appendix
\section{}
\setcounter{table}{0}\renewcommand\thetable{\Alph{section}.\arabic{table}}
\caption{Lagrangian radii $63.2\%$, r$_{63.2}$, in AU, final mean planetary distance, $\stretchleftright{<}{r}{>}$, in AU, final mean planetary eccentricity, $\stretchleftright{<}{e}{>}$, and final planetary mass, M$_{p}$, in M$_{Jup}$, of the model of the stellar perturber of mass 0.5 M$_{\odot}$.}
\begin{tabular}{ c|c|c|c|c }
\cline{1-5}
\hline
\multicolumn{5}{ c }{r$_{63.2}$}\\
\hline \hline
\multicolumn{1}{ c|}{ r$_{p}$} &
\multicolumn{2}{c|}{$0^{\circ}$} &
\multicolumn{2}{c }{$180^{\circ}$} \\
\hline
\hline
& 
{ {time$_0$}} &
{ {time$_{end}$}} &
{ {time$_0$}} &
{ {time$_{end}$}}
\\
\hline
{100} &
{ {38.59}} & 
{ {44.57}} & 
{ {39.12}} & 
{ {36.24}}
\\
\hline
{200} & 
{ {38.72}} & 
{ {46.66}} & 
{ {38.81}} & 
{ {61.02}}
\\
\hline
{300 } &
{ {38.85}} & 
{ {45.95}} & 
{ {38.82}} & 
{ {60.14}}
\\
\hline
{400 } & 
{ {39.37}} & 
{ {50.07}} &
{ {38.02}} & 
{ {48.23}}
\\
\hline
{500 } & 
{ {38.85}} & 
{ {47.38}} & 
{ {38.70}} & 
{ {46.79}}
\\
\hline
\end{tabular}

\vspace{0.3cm}

\centering
\begin{tabular}{ c|c|r|c|c|c }
\hline
\multicolumn{6}{ c }{$0^{\circ}$} \\
\hline
{ r$_{p}$} & 100 & 200 & 300 &  400&  500 \\
\hline
$\stretchleftright{<}{r}{>}$ 
                    & { {5.06}}
                    & { {5.11}}
                    & { {6.04}}
                    & { {5.90}}
                    & { {5.68}} \\
\hline
$\stretchleftright{<}{e}{>}$ & $ {0.05}$
	            & $ {0.02}$ 
	            & $ {0.02}$
	            & $ {0.11}$
	            &  $ {0.01}$ \\
\hline
M$_{p}$  & $ {1.22}$ 
	                    &  $ {1.22}$ 
	                    & $ {1.19}$
	                    & $ {1.20}$
	                    & $ {1.20}$ \\
	\hline
\end{tabular}

\vspace{0.3cm}

\centering
\begin{tabular}{ c|c|r|c|c|c }
\hline
\multicolumn{6}{ c }{$180^{\circ}$} \\
\hline
{ r$_{p}$} & 100 & 200 & 300 &  400&  500 \\
\hline
$\stretchleftright{<}{r}{>}$ 
                    & { {7.93}}
                    & { {5.84}}
                    & { {6.16}}
                    & { {6.50}}
                    & { {6.13}} \\
\hline
$\stretchleftright{<}{e}{>}$ & $ {0.20}$ 
	               & $ {0.03}$ 
	               & $ {0.08}$  
	               & $ {0.07}$
	               & $ {0.06}$\\
\hline
M$_{p}$  & $ {1.13}$ 
	                       & $ {1.17}$  
	                       & $ {1.19}$ 
	                       & $ {1.17}$
	                       & $ {1.18}$\\
	\hline
\end{tabular}
\end{table*}

\newpage
\begin{table*}
\centering
\setcounter{table}{1}\renewcommand\thetable{\Alph{section}.\arabic{table}}
\caption{As in Table A.1 but for a stellar perturber of mass 0.8 M$_{\odot}$. }
\begin{tabular}{ c|c|c|c|c }
\cline{1-5}
\hline
\multicolumn{5}{ c }{r$_{63.2}$}\\
\hline \hline
\multicolumn{1}{ c|}{ r$_{p}$} &
\multicolumn{2}{c|}{$0^{\circ}$} &
\multicolumn{2}{c }{$180^{\circ}$} \\
\hline
\hline
& 
{ {time$_0$}} &
{ {time$_{end}$}} &
{ {time$_0$}} &
{ {time$_{end}$}}
\\
\hline
{100 } & 
{ {38.50}} & 
{ {42.64}} & 
{ {39.07}} & 
{ {45.75}}
\\
\hline
{200 } & 
{ {38.59}} & 
{ {45.15}} & 
{ {38.24}} & 
{ {49.26}} 
\\
\hline
{300 } & 
{ {39.04}} & 
{ {47.53}} & 
{ {38.16}} & 
{ {46.12}}
\\
\hline
{400 } & 
{ {38.90}} & 
{ {48.68}} & 
{ {38.69}} & 
{ {62.48}}
\\
\hline
{500 } & 
{ {38.68}} & 
{ {45.64}} & 
{ {38.21}} & 
{ {45.81}}
\\
\hline
\end{tabular}

\vspace{0.5cm}

\centering
\begin{tabular}{ c|c|r|c|c|c }
\hline
\multicolumn{6}{ c }{$0^{\circ}$} \\
\hline
{ r$_{p}$} & 100 & 200 & 300 &  400&  500 \\
\hline
$\stretchleftright{<}{r}{>}$
                    & { {6.65}} 
                    & { {6.41}} 
                    & { {6.00}}
                    & { {5.78}}
                    & { {5.53}} \\
\hline
$\stretchleftright{<}{e}{>}$ & $ {0.13}$
	            & $ {0.06}$ 
	            & $ {0.11}$
	            & $ {0.11}$
	            &  $ {0.09}$ \\
\hline
M$_{p}$  & $ {1.17}$ 
	                    & $ {1.18}$ 
	                    & $ {1.18}$
	                    & $ {1.20}$
	                    & $ {1.22}$ \\
\hline
\end{tabular}

\vspace{0.5cm}

\centering
\begin{tabular}{ c|c|r|c|c|c }
\hline
\multicolumn{6}{ c }{$180^{\circ}$} \\
\hline
{ r$_{p}$} & 100 & 200 & 300 &  400&  500 \\
\hline
$\stretchleftright{<}{r}{>}$ 
                    & { {7.73}}
                    & { {5.66}} 
                    & { {6.58}}
                    & { {6.36}}
                    & { {6.18}} \\

\hline
$\stretchleftright{<}{e}{>}$ & $ {0.28}$
	            & $ {0.12}$ 
	            & $ {0.09}$
	            & $ {0.16}$
	            &  $ {0.12}$ \\
\hline
M$_{p}$  & $ {1.12}$ 
	                    &  $ {1.20}$ 
	                    & $ {1.16}$
	                    & $ {1.12}$
	                    & $ {1.18}$ \\
	\hline
\end{tabular}
\end{table*}

\newpage
\centering
\begin{table*}
\setcounter{table}{2}\renewcommand\thetable{\Alph{section}.\arabic{table}}
\caption{As in Table A.1 but for a stellar perturber of mass 1 M$_{\odot}$.}
\centering
\begin{tabular}{ c|c|c|c|c }
\cline{1-5}
\hline
\multicolumn{5}{ c }{r$_{63.2}$}\\
\hline \hline
\multicolumn{1}{ c|}{ r$_{p}$} &
\multicolumn{2}{c|}{$0^{\circ}$} &
\multicolumn{2}{c }{$180^{\circ}$} \\
\hline
\hline
& 
{ {time$_0$}} &
{ {time$_{end}$}} &
{ {time$_0$}} &
{ {time$_{end}$}}
\\
\hline
{100 } & 
{ {39.05}} & 
{ {40.63}} & 
{ {39.06}} & 
{ {27.82}}
\\
\hline
{200 } & 
{ {39.03}} & 
{ {44.71}} & 
{ {38.07}} & 
{ {41.21}}
\\
\hline
{300 } & 
{ {38.54}} & 
{ {47.40}} & 
{ {39.14}} & 
{ {44.09}}
\\
\hline
{400 } & 
{ {38.84}} & 
{ {48.28}} & 
{ {38.72}} & 
{ {62.26}}
\\
\hline
{500 } & 
{ {38.83}} & 
{ {52.32}} & 
{ {39.23}} & 
{ {58.73}}
\\
\hline
\end{tabular}

\vspace{0.3cm}

\centering
\begin{tabular}{ c|c|c|c|c|c }
\hline
\multicolumn{6}{ c }{ {$0^{\circ}$}} \\
\hline
{ r$_{p}$} & 100 & 200 & 300 &  400&  500 \\
\hline
$\stretchleftright{<}{r}{>}$ 
                    &  {6.74}
                    &  {5.89}
                    &  {6.16}
                    &  {5.85}
                    &  {7.55} \\
\hline
$\stretchleftright{<}{e}{>}$ & $ {0.05}$ 
	               &$ {0.23}$  
	               & $ {0.08}$
	               & $ {0.31}$
	               & $ {0.24}$\\
\hline
M$_{p}$  & $ {1.14}$ 
	                   & $ {1.22}$ 
	                   & $ {1.20}$ 
	                   & $ {1.23}$ 
	                   & $ {1.17}$ \\
	\hline
\end{tabular}

\vspace{0.3cm}
\centering
\begin{tabular}{ c|c|c|c|c|c }
\hline
\multicolumn{6}{ c }{$180^{\circ}$} \\
\hline
{ r$_{p}$} & 100 & 200 & 300 &  400&  500 \\
\hline
$\stretchleftright{<}{r}{>}$ 
                    &  {6.06}
                    &  {5.70}  
                    &  {7.02}
                    &  {5.96}
                    &  {6.12} \\
\hline
$\stretchleftright{<}{e}{>}$ & $ {0.18}$ 
	               & $ {0.10}$ 
	               & $ {0.04}$
	               & $ {0.06}$
	               & $ {0.09}$\\
\hline
M$_{p}$  & $ {1.20}$ 
	                       & $ {1.21}$  
	                       & $ {1.16}$ 
	                       & $ {1.18}$ 
	                       & $ {1.22}$ \\
\hline
\end{tabular}
\end{table*}
\newpage

\begin{table*}
\centering
\setcounter{table}{3}\renewcommand\thetable{\Alph{section}.\arabic{table}}
\caption{As in Table A.1 but for a stellar perturber of mass 3 M$_{\odot}$.}
\begin{tabular}{ c|c|c|c|c }
\cline{1-5}
\hline
\multicolumn{5}{ c }{r$_{63.2}$}\\
\hline \hline
\multicolumn{1}{ c|}{ r$_{p}$} &
\multicolumn{2}{c|}{$0^{\circ}$} &
\multicolumn{2}{c }{$180^{\circ}$} \\
\hline
\hline
& 
{ {time$_0$}} &
{ {time$_{end}$}} &
{ {time$_0$}} &
{ {time$_{end}$}}
\\
\hline
{100 } & 
{ {38.23}} &
{ {39.37}} &
{ {39.36}} & 
{ {26.11}}
\\
\hline
{200 } &
{ {38.22}} & 
{ {48.23}}  &
{ {38.64}} & 
{ {35.14}}
\\
\hline
{300 } & 
{ {38.27}} &
{ {48.73}} & 
{ {39.11}} &
{ {55.12}}
\\
\hline
{400 } &
{ {38.43}} &
{ {46.28}} &
{ {39.25}} &
{ {54.58}}
\\
\hline
{500 } & 
{ {38.39}} &
{ {53.27}} &
{ {38.80}} &
{ {51.96}}
\\
\hline
\end{tabular}

\vspace{0.3cm}

\centering
\begin{tabular}{ c|c|c|c|c|c }
\hline
\multicolumn{6}{ c }{$0^{\circ}$} \\
\hline
{ r$_{p}$} & 100 & 200 & 300 &  400&  500 \\
\hline
$\stretchleftright{<}{r}{>}$ 
                    & { {5.55}}
                    & { {6.05}}
                    & { {4.61}}
                    & { {5.82}}
                    & { {5.40}} \\
\hline
$\stretchleftright{<}{e}{>}$ & $ {0.06}$
	            & $ {0.42}$ 
	            & $ {0.27}$
	            & $ {0.12}$
	            &  $ {0.12}$ \\
\hline
M$_{p}$  & $ {1.19}$ 
	               & $ {1.21}$ 
	               & $ {1.30}$
	               & $ {1.21}$
	               & $ {1.22}$ \\
	\hline
\end{tabular}

\vspace{0.3cm}

\begin{tabular}{ c|c|c|c|c|c }
\hline
\multicolumn{6}{ c }{$180^{\circ}$} \\
\hline
{ r$_{p}$} & 100 & 200 & 300 &  400&  500 \\
\hline
$\stretchleftright{<}{r}{>}$ 
                    & { {5.61}}
                    & { {7.62}}
                    & { {5.66}}
                    & { {6.12}}
                    & { {6.23}} \\
\hline
$\stretchleftright{<}{e}{>}$ & $ {0.10}$
	            & $ {0.62}$ 
	            & $ {0.03}$
	            & $ {0.12}$
	            &  $ {0.18}$ \\
\hline
M$_{p}$  & $ {1.20}$ 
	                    & $ {1.17}$ 
	                    & $ {1.24}$
	                    & $ {1.21}$
	                    & $ {1.20}$ \\
\hline
\end{tabular}
\end{table*}
\newpage
\begin{table*}
\setcounter{table}{4}\renewcommand\thetable{\Alph{section}.\arabic{table}}
\caption{As in Table A.1 but for a stellar perturber of mass 8 M$_{\odot}$.}
\centering
\begin{tabular}{ c|c|c|c|c }
\cline{1-5}
\hline
\multicolumn{5}{ c }{r$_{63.2}$}\\
\hline \hline
\multicolumn{1}{ c|}{ r$_{p}$} &
\multicolumn{2}{c|}{$0^{\circ}$} &
\multicolumn{2}{c }{$180^{\circ}$} \\
\hline
\hline
& 
{ {time$_0$}} &
{ {time$_{end}$}} &
{ {time$_0$}} &
{ {time$_{end}$}}
\\
\hline
{100 } & 
{ {38.62}} &
{ {47.55}} & 
{ {39.04}} &
{ {21.83}}
\\
\hline
{200 } & 
{ {38.64}} &
{ {66.35}} & 
{ {38.81}} &
{ {30.20}}
\\
\hline
{300} & 
{ {39.14}} &
{ {60.59}}  & 
{ {38.47}} &
{ {39.98}}
\\
\hline
{400 } & 
{ {38.88}} &
{ {53.30}} & 
{ {39.11}} &
{ {45.61}}
\\
\hline
{500 } & 
{ {38.76}} &
{ {52.87}} & 
{ {38.82}} &
{ {48.25}}
\\
\hline
\end{tabular}

\vspace{0.3cm}

\centering
\begin{tabular}{ c|c|r|c|c|c }
\hline
\multicolumn{6}{ c }{$0^{\circ}$} \\
\hline
{ r$_{p}$} & 100 & 200 & 300 &  400&  500 \\
\hline
$\stretchleftright{<}{r}{>}$ 
                    & { {7.42}}
                    & { {7.06}}
                    & { {5.97}}
                    & { {5.55}}
                    & { {5.93}} \\
\hline
$\stretchleftright{<}{e}{>}$ & $ {0.27}$
	            & $ {0.42}$ 
	            & $ {0.24}$
	            & $ {0.46}$
	            &  $ {0.24}$ \\
\hline
M$_{p}$  & $ {1.18}$ 
	                    & $ {1.20}$ 
	                    & $ {1.24}$
	                    & $ {1.27}$
	                    & $ {1.23}$ \\
\hline
\end{tabular}

\vspace{0.3cm}

\begin{tabular}{ c|c|r|c|c|c }
\hline
\multicolumn{6}{ c }{$180^{\circ}$} \\
\hline
{ r$_{p}$} & 100 & 200 & 300 &  400&  500 \\
\hline
$\stretchleftright{<}{r}{>}$ 
                    & { {5.00}}
                    & { {5.59}}
                    & { {6.67}}
                    & { {7.14}}
                    & { {4.72}} \\
\hline
$\stretchleftright{<}{e}{>}$ & $ {0.13}$
	            & $ {0.09}$ 
	            & $ {0.53}$
	            & $ {0.08}$
	            &  $ {0.09}$ \\
\hline
M$_{p}$  & $ {1.26}$ 
	                    & $ {1.23}$ 
	                    & $ {1.21}$
	                    & $ {1.17}$
	                    & $ {1.28}$ \\
\hline
\end{tabular}
\end{table*}

\label{lastpage}
\end{document}